\documentclass[epj]{webofc}
\usepackage[varg]{txfonts}   
\pdfoutput=1 
\wocname{}
\woctitle{}

\usepackage{pifont}
\usepackage{enumitem}
\usepackage{longtable}

\RequirePackage{lineno}
\usepackage{rotating}
\usepackage{graphicx}
\usepackage{graphics}

\usepackage{subfigure}
\usepackage{wrapfig}
\usepackage{dcolumn}
\usepackage{bm}
\usepackage{longtable}
\usepackage[percent]{overpic}
\usepackage{color}

\usepackage{float}
\usepackage{psfrag}

\def\pt{p_{\rm T}}
\def\sNN{\mbox{$\sqrt{s_{_{\rm NN}}}$}}   
\def\av#1{\langle #1 \rangle}

\def\sNN{\mbox{$\sqrt{s_{_{\rm NN}}}$}}

\begin{document}
\selectlanguage{english}
\title{Classifiers for centrality determination in proton-nucleus and nucleus-nucleus collisions}
%
%

\author{ Igor  Altsybeev\inst{1}\fnsep\thanks{\email{
i.altsybeev@spbu.ru 
}} 
and 
{ Vladimir Kovalenko\inst{1}\fnsep\thanks{\email{v.kovalenko@spbu.ru 
}}}
}

\institute{Saint-Petersburg State University,
7/9 Universitetskaya nab., St. Petersburg, 199034 Russia}

\abstract{
Centrality, as a geometrical property of the collision, is crucial for the physical interpretation of nucleus-nucleus and proton-nucleus experimental data. However, it cannot be directly accessed in event-by-event data analysis. Common methods for centrality estimation in A-A and p-A collisions usually rely on a single detector (either on the signal in zero-degree calorimeters or on the multiplicity in some semi-central rapidity range). In the present work, we made an attempt to develop an approach for centrality determination that is based on machine-learning techniques and utilizes information from several detector subsystems simultaneously. Different event classifiers are suggested and evaluated for their selectivity power in terms of the number of nucleons-participants and the impact parameter of the collision. Finer centrality resolution may allow to reduce impact from so-called volume fluctuations on physical observables being studied in heavy-ion experiments like ALICE at the LHC and fixed target experiment NA61/SHINE on SPS.
}
\maketitle

\section{Introduction}

Machine-learning (ML) techniques have been used in High-Energy Physics (HEP) 
so far
in a limited number of ways.
The most common application of the ML is a search for rare processes 
(for example, boosted decision trees were used in searches
for $B^0_s\rightarrow\mu^+\mu^-$ decay \cite{decay_B0s}).
Another use case is a trigger optimization \cite{topologica_trigger_opt}.
In experiments dedicated to studies of heavy-ion collisions, like ALICE at the LHC,
 ML is used mainly for detector response optimization,
as described in overview \cite{MFloris_data_science_in_ALICE}.
For example,
the Bayesian approach is used 
to more effectively combine the particle identification capabilities of various detectors
\cite{Bayesian_PID}.

It is interesting to find new applications of ML 
for physics analysis  in experiments of heavy-ion collisions. 
In the present work, we tried to address the task of the centrality determination.
The centrality is a key parameter in the study of the properties of QCD matter at extreme temperature and energy density, because it is directly related 
to the initial overlap region of the colliding nuclei~\cite{ALICE_centrality_PbPb}.
The centrality is usually expressed as a percentage of the total nuclear interaction cross section $\sigma$:
$c=\int_0^b \frac{d\sigma}{dx}dx$, where
impact parameter ($b$) is the distance between the centers of the two colliding nuclei.
Taking the full impact parameter range 
as 0-100\%, 
peripheral collisions have centrality closer to 100\%,
while the most central events are close to 0\%.

Impact parameter $b$ is not directly accessible in experiment.
In order to obtain estimation of centrality  experimentally,
a signal distribution in some detecting system is usually used:
the distribution is divided into {\it centrality classes},
which are then related  to some interval of the impact parameter, with some estimated resolution.
To make this relation, the geometrical Glauber model \cite{Glauber} is usually used,
which treats a nuclear collision as a superposition of binary nucleon-nucleon interactions.
A nucleon that undergoes one or more  collisions with nucleons of the other nucleus,
is usually called a {\it participant nucleon}.
The volume of the initial overlap region is expressed via the number of participant nucleons
$N_{\rm part}$. 
A number of binary nucleon-nucleon collisions is usually denoted as $N_{\rm coll}$.
A number of {\it spectator nucleons} $N_{\rm spec}$  constitutes the part of the nuclear volume
not involved in the interaction. 

Any increase in centrality resolution 
would be beneficial
for certain types of physics analysis.
For example, higher centrality resolution is desirable 
in class 0-1\% of Pb-Pb collisions
to select the most central events,
where double-peaked structure was observed
in azimuthal profile of two-particle correlation function
\cite{harmonic_decomposition}.

The present exploratory work is dedicated to
ML-based task for centrality determination. 
The aim is to select ``centrality classes''
with improved resolution in terms of the impact parameter $b$
in Pb-Pb  and $N_{\rm part}$
in p-Pb collisions 
using signals from several  subsystems of the detector simultaneously.
The task is performed using simulated events.
Dealing with
detector-induced losses in efficiency, contamination by secondary particles
and application of this technique to real data   
are out of the scope of this article.


\section{Centrality determination in ALICE experiment}

\begin{figure}[b]
\centering
\subfigure[a][]
{
\begin{overpic}[width=0.51\textwidth, scale=1, bb=0 0 1105 655]
{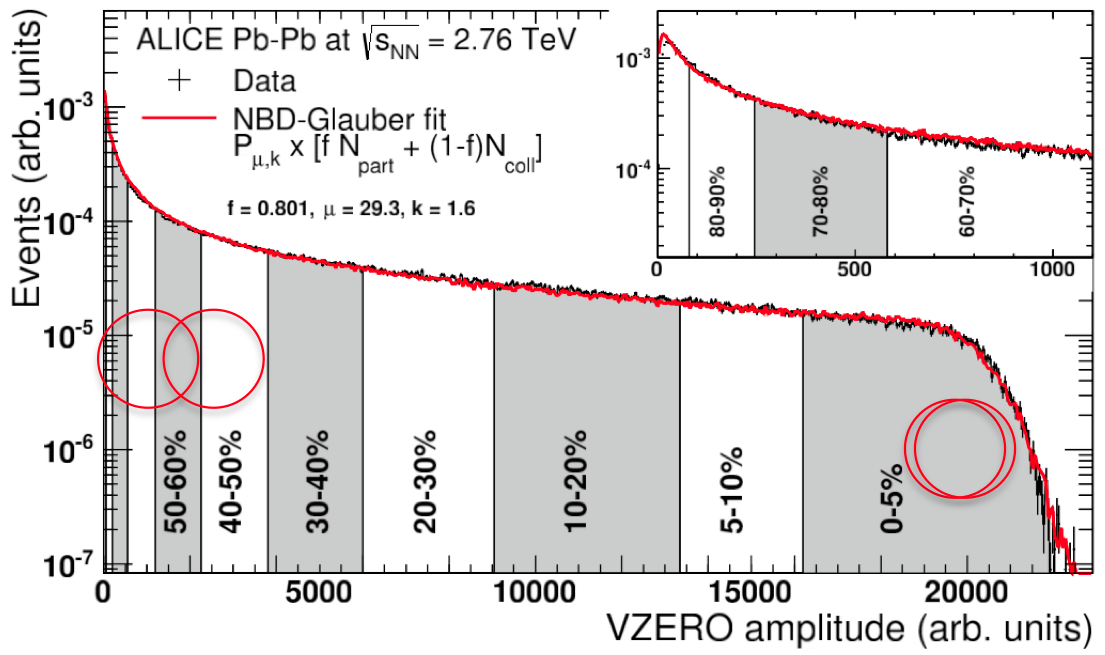}
\end{overpic}
}
\hspace{-0.1cm}
\subfigure[a][]
{
\begin{overpic}[width=0.44\textwidth, bb=0 0 920 727]
{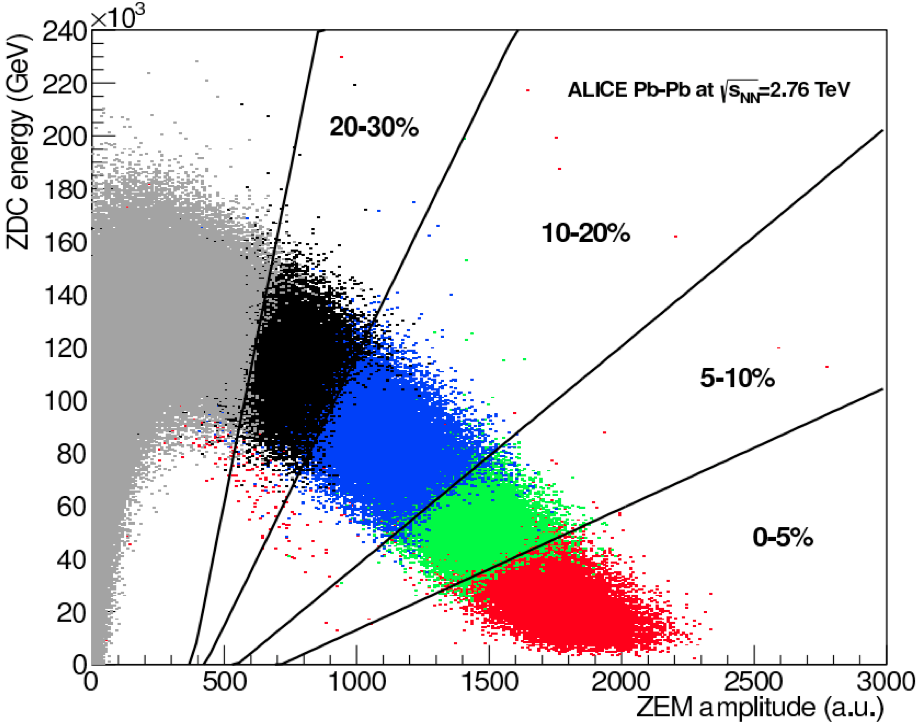}
\end{overpic}
}
\caption{
(a): Distribution of the sum of amplitudes in the VZERO scintillators (V0M). The distribution
is fitted with the NBD-Glauber fit shown as a line, 
obtained centrality classes are indicated.
(b): 
Spectator energy deposited in the ZDCs  as a function of ZEM amplitude
\cite{ALICE_centrality_PbPb}.
}
\label{fig:V0M_ZEMvsZDC}
\end{figure}

In the ALICE experimental setup \cite{ALICE_performance},
the most commonly used detecting system for centrality determination is
a pair of so-called VZERO detectors,
two forward scintillator arrays
with coverage $-3.7<\eta<-1.7$ (VZERO-C) and $2.8<\eta<5.1$ (VZERO-A).
Since these 
are non-tracking detectors, 
signal in each VZERO is proportional to charged particle multiplicity
of primary tracks and secondary tracks from weak decays and detector material.
The summed signal in both VZERO scintillators
is usually denoted as V0M. 
Distribution of the V0M amplitude shown in figure \ref{fig:V0M_ZEMvsZDC} (a)
is fitted with the Glauber model coupled with 
 negative binomial distribution (NBD),
and divided into centrality classes \cite{ALICE_centrality_PbPb}.
Values of $N_{\rm part}$ and $N_{\rm coll}$ 
in each centrality class 
are extracted from the fit.


Another set of detectors used
for centrality determination
are the Zero-Degree Calorimeters (ZDC) placed
at $\approx\pm 110$ m in the LHC tunnel.
The energy deposited in the ZDCs is directly
related to the number of spectator nucleons $N_{\rm spec}$.
Combining signal from ZDCs with the energy measured by small EM calorimeters
(ZEM) placed at A-side of ALICE
at  $4.8 < \eta < 5.7$,
a two-dimensional distribution can be plotted (figure \ref{fig:V0M_ZEMvsZDC}, b).
The centrality classes 
are determined from this plot by splitting it with lines,
which are assumed  
to intersect at some common point \cite{ALICE_centrality_PbPb}.

Charged primary particles are reconstructed in ALICE with the central barrel detectors combining information from the Inner Tracking System (ITS) and the Time Projection Chamber (TPC). 
The TPC, together with the ITS, provides charged particle momentum measurement, particle identification and vertex determination.
Both detectors are located inside the 0.5~T solenoidal field
and have full azimuthal coverage for track reconstruction 
within a pseudo-rapidity window of $|\eta| < 0.8$.

\section{ML task for simulated Pb-Pb events in the conditions of ALICE experiment}
\label{sec:ML_task_PbPb}

\begin{figure}[h]
\centering
\begin{overpic}[width=0.85\textwidth, scale=1, bb=0 0 1378 312]
{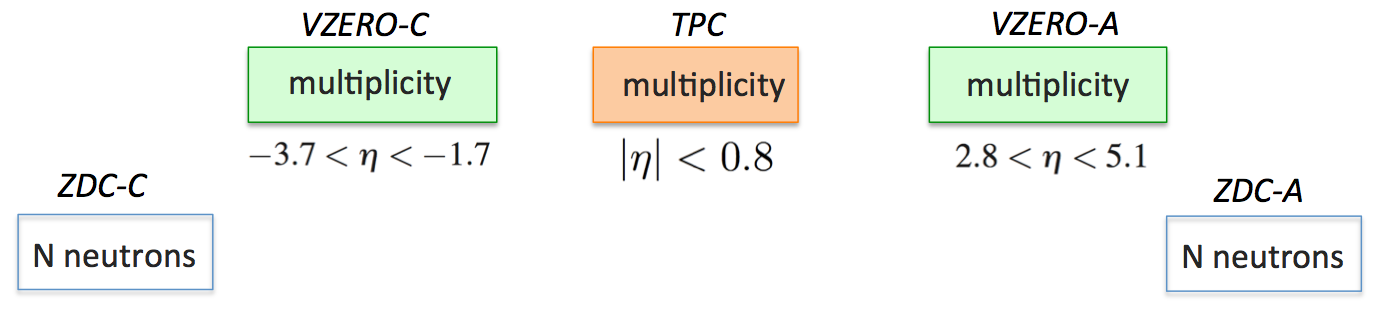}
\end{overpic}
\caption{
Detecting systems, from which five features for the ML task were taken:
multiplicities in acceptances of  VZERO-A, C, TPC,
and numbers of nucleons-spectators in each of two ZDCs.
}
\label{fig:features_in_ALICE}
\end{figure}

The AMPT Monte Carlo
event generator \cite{AMPT}  was used to simulate Pb-Pb collisions at $\sNN=2.76$~TeV. 
The simulation contains {\it no} detector response,
so ``signals'' of the real detector were emulated
directly by taking generated particles in certain areas of the phase space.
Five main features were selected for the ML task
 in correspondence with the subsystems of the ALICE detector (figure \ref{fig:features_in_ALICE}).
These are multiplicities of charged particles obtained in AMPT simulation
within acceptances of 
each of the VZERO scintillators (A, C)  and the TPC, 
and also numbers of neutrons-spectators,
which in the ``ideal world" would be measured in each of the ZDCs (A and C).
Also, three additional features were introduced 
in order to study their importance 
for improvement of the final centrality resolution: 
average transverse momentum of particles in the TPC within each event ($\av\pt$),
kaon-to-pion  (K$/\pi$) and 
proton-to-pion (p$/\pi$) ratios  in each event.
These observables are known to be dependent on centrality 
of heavy-ion collisions
and possibly possess some discriminative power.

\subsection{Regression task}

It is interesting 
to find out whether we have
some improvement 
in centrality resolution 
if, instead of V0M,
 we perform event selection for centrality classes
using ML regressor.
To perform ML regression task for centrality determination,
TMVA package \cite{TMVA2007}  of version 4.2.0 was used.
For this task,
 a feature set of five variables, described above, was taken,
with the impact parameter $b$ 
as a target for regression.
Among several probed regressors, good performance was obtained,
in particular, with a Boosted Decision Tree with gradient boosting (denoted as BDTG),
results from which 
are used here for illustration.
Booking in TMVA was done with the following line: \\
 \texttt{\hspace*{0.27cm} \footnotesize 
        factory->BookMethod( TMVA::Types::kBDT, "BDTG", 
                             "!H:!V:NTrees=2000::BoostType=Grad: \\
\hspace*{0.27cm}
Shrinkage=0.1:UseBaggedBoost:BaggedSampleFraction=0.5:nCuts=20:MaxDepth=3:MaxDepth=4");
} \\
To focus on more central collisions,
events for the analysis were preselected 
using V0M estimator to be within 0-10\%  centrality class 
(see centrality classes in figure \ref{fig:V0M_ZEMvsZDC}, a).
This preselected set contains 400k of events,
and it was randomly split into two halves: 
200k events for training 
and 200k for testing.

\begin{figure}[H]
\centering
\sidecaption
\begin{overpic}[width=0.42\textwidth, bb=0 0 863 667, clip=true, trim=0 0 5 5]
{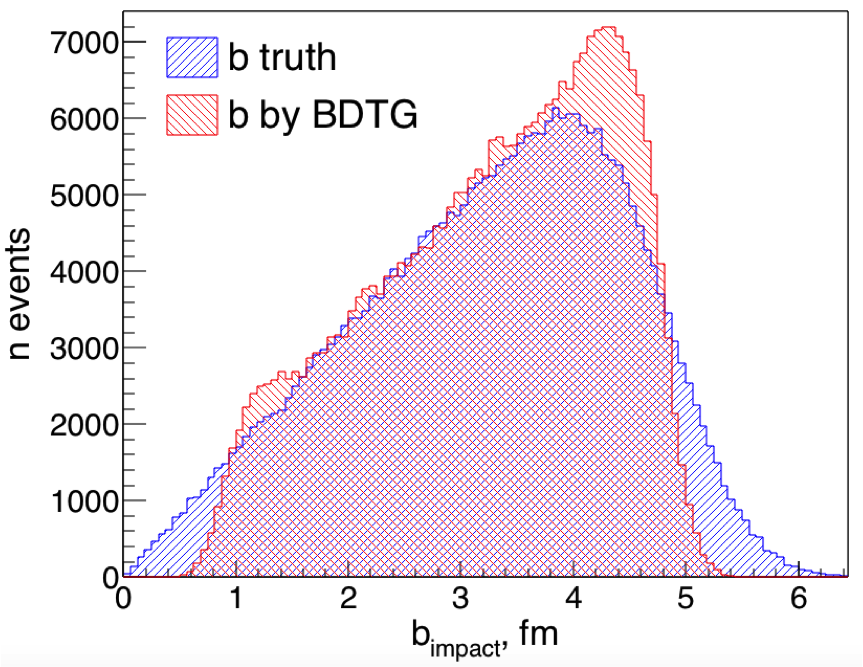}
\end{overpic}
\caption{
Distribution of impact parameter $b$ for events in 0-10\% centrality class  
preselected by simulated V0M signal  (wide distribution, shown in blue).
Output from BDTG regressor with impact parameter $b$ as a target is narrower and is shown in red.
}
\label{fig:b_regression_PbPb}
\end{figure}

Distribution of the {\it truth} impact parameter $b$ in events, preselected by 
V0M,
is shown in 
figure \ref{fig:b_regression_PbPb} 
as a blue shaded area.
Output from BDTG regressor, which is a distribution of {\it estimated} values of $b$,
 is shown on the same plot in red. 
It is ``squeezed" from both left and right sides, demonstrating inability of
the regressor 
to reach left and right extremes of the truth distribution. 

\begin{figure}[b]
\centering
\subfigure[a][]
{
\begin{overpic}[width=0.33\textwidth, bb=0 0 798 575]
{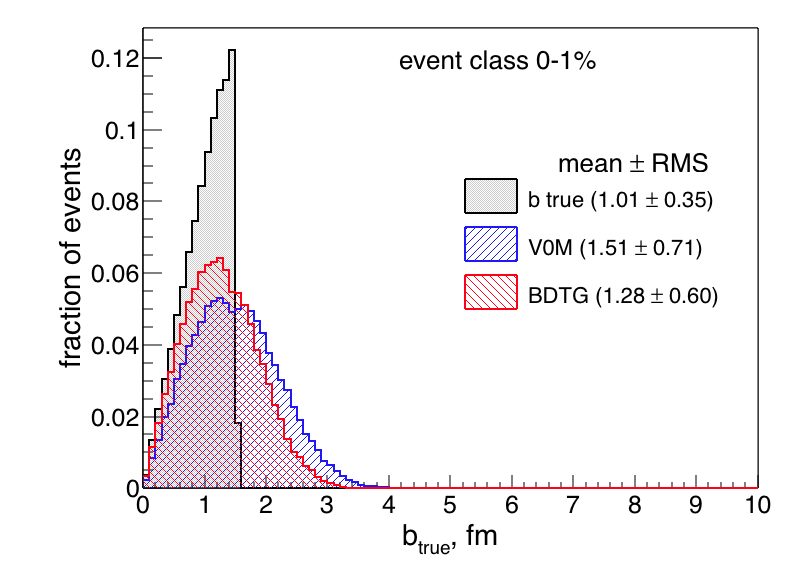}
\end{overpic}
}
\hspace{-0.5cm}
\subfigure[a][]
{
\begin{overpic}[width=0.33\textwidth, bb=0 0 798 575]
{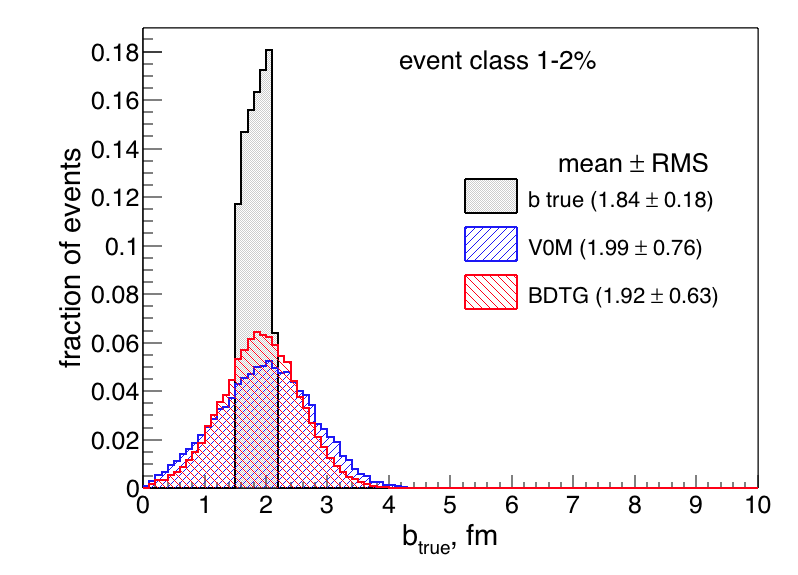}
\end{overpic}
}
\hspace{-0.5cm}
\subfigure[a][]
{
\begin{overpic}[width=0.33\textwidth, bb=0 0 798 575]
{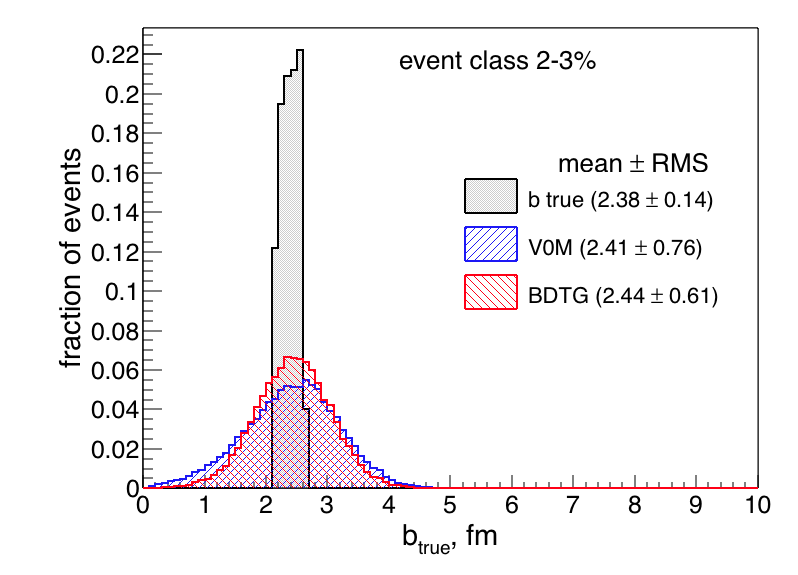}
\end{overpic}
}
\caption{
Distributions of impact parameter within centrality classes 0-1, 1-2 and 2-3\%
selected by V0M and BDTG estimators
(color code is different from figure \ref{fig:b_regression_PbPb}).
For comparison, grey distributions
show slices of 1\% width 
when slicing into centrality classes is done by truth $b$ itself.
}
\label{fig:distr_bTrue_V0M_BDTG}
\end{figure}

In order to use BDTG regressor
as another (alternative) centrality estimator, 
its output was 
split into ``centrality bins''
in the same manner as it is done for conventional V0M distribution
(recall figure~\ref{fig:V0M_ZEMvsZDC},~a).
To compare performance of the new BDTG estimator
with the V0M in terms of impact parameter resolution,
 outputs from both estimators 
were split into 10 centrality bins (this was done for 0-10\% preselected by V0M events,
so obtained classes are of 1\% width).
Figure \ref{fig:distr_bTrue_V0M_BDTG}
shows {\it truth} impact parameter distributions
for events 
within 0-1\%, 1-2\% and 2-3\% centrality classes
selected by V0M (in blue) and selected using new BDTG estimator
(in red). 
It can be seen that 
when selection is done with the BDTG estimator,
distributions of $b$ are narrower 
than with V0M selection, 
which means higher resolution in terms of $b$.
Also, in most central 0-1\% event class selected with the BDTG estimator,
the $b$-distribution goes closer towards $b$=0
(figure \ref{fig:distr_bTrue_V0M_BDTG}, a), indicating that within this 
class we indeed select more central events than with V0M.

\begin{figure}[H]
\centering
\sidecaption
\begin{overpic}[width=0.48\textwidth]
{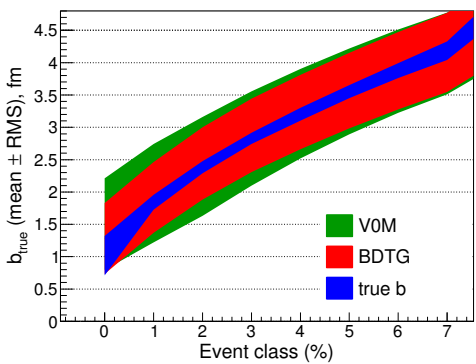}
\end{overpic}
\caption{
Mean $\pm$ RMS of impact parameter $b$ distributions
for events within centrality classes of 1\% width,
which are selected using 
simulated multiplicity distribution in V0M acceptance (in green)
and BDTG (in red) estimators.
Blue area covers 
mean $\pm$ RMS  of  $b$ distributions,
when slicing into centrality classes is done by truth $b$ itself.
}
\label{fig:b_vs_eventClass_mean_rms}
\end{figure}

The same information is summarized in figure \ref{fig:b_vs_eventClass_mean_rms},
where each colored region 
covers the mean values of $b$ distributions $\pm$ their RMS
in  centrality classes of 1\%-width in 0-10\% range: 
the red area stands for centrality selection by the BDTG estimator, and
it is narrower than the green area for selection using V0M, 
especially for the most central events.
For comparison, 
the blue area 
is for the case when
slicing into centrality classes is done by truth $b$ itself.

\subsection{Classification task for the most central events}

In this section,
centrality determination task 
is considered as a classification problem.
Namely, within 0-10\% class of events, preselected with 
simulated multiplicity distribution in VZERO acceptance,
the  most central  0-1\% events (with $b_{\rm impact} < 1.5$ fm)
are considered as a signal, all other events -- as a background.
Purity is defined as a fraction
of events, selected by classifier, for which 
impact parameter is indeed $<1.5$~fm.
The challenge is to increase the purity of 0-1\% selected events.
The same set of five descriminating features was used.

\begin{figure}[t]
\centering
\subfigure[a][]
{
\begin{overpic}[width=0.48\textwidth, bb=0 0 1045 728]
{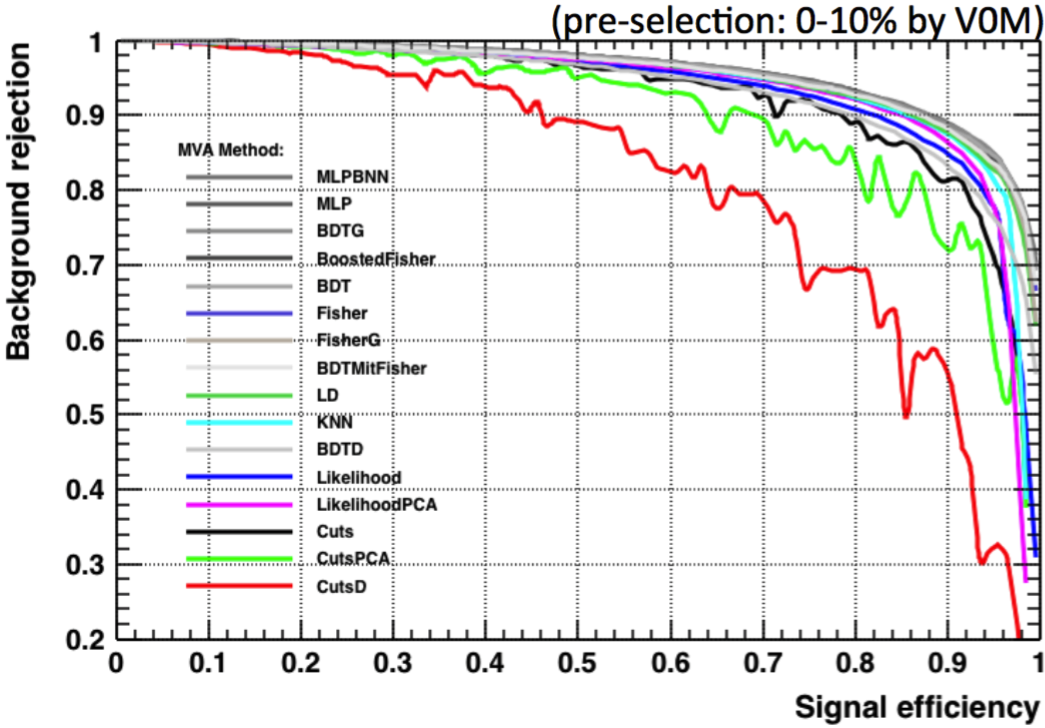}
\end{overpic}
}
\hspace{-0.2cm}
\subfigure[a][]
{
\begin{overpic}[width=0.49\textwidth]
{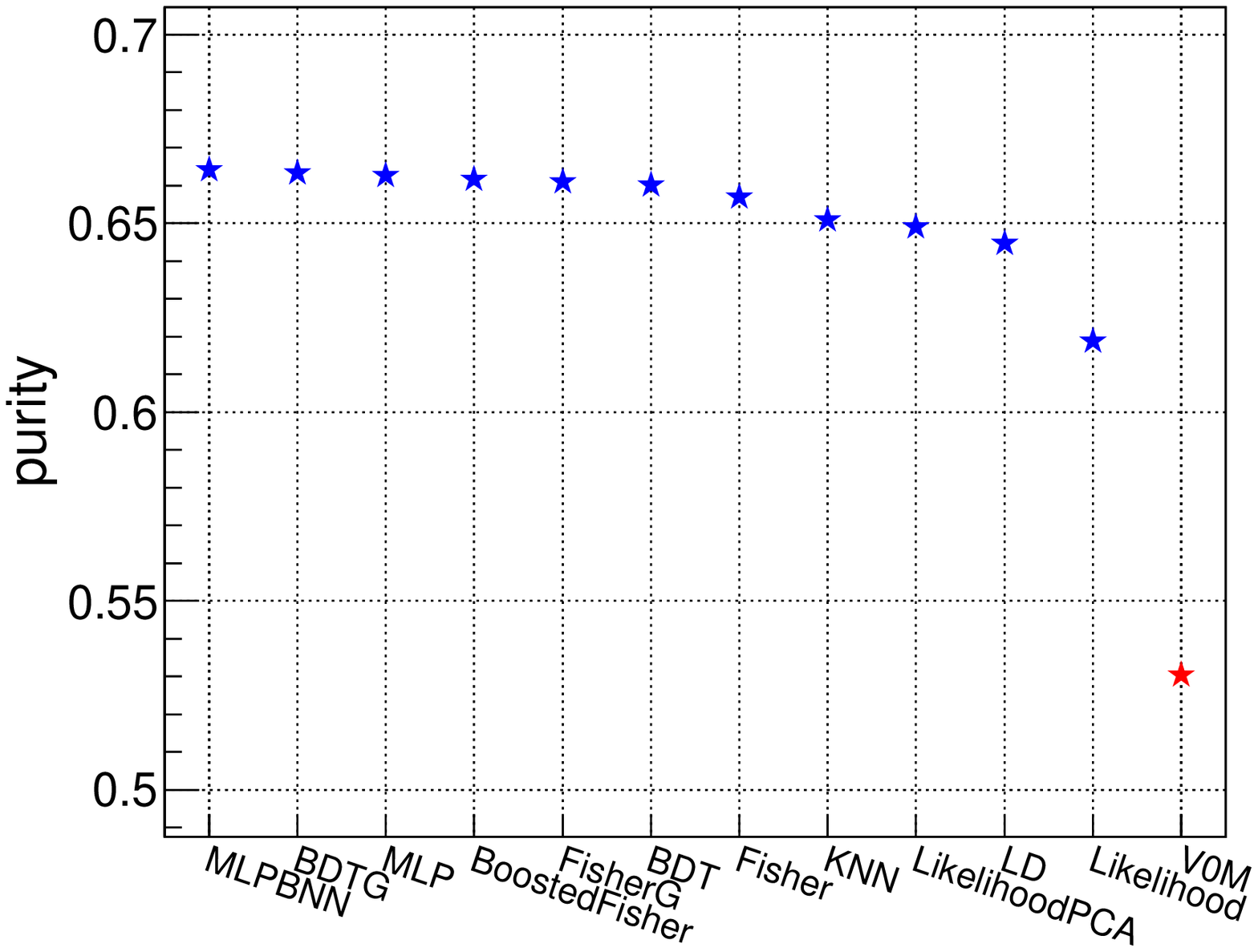}
\end{overpic}
}
\caption{
(a): Signal efficiency $vs$ background rejection for a set of classifiers (TMVA).
(b): Purities reached with classifiers in 0-1\% centrality class
in terms of impact parameter. 
Red star is for purity of  0-1\% event class with V0M selection.
}
\label{fig:classification_PbPb}
\end{figure}

Figure \ref{fig:classification_PbPb} (a) 
shows signal efficiency $vs$ background rejection for a large set of classifiers, probed using TMVA package.
It can be seen, that, except for simple cut-based classifiers,
all estimators provide similar performance, with slightly 
better results for neural networks (MLP) and boosted decision trees.
Pad (b) of the same figure presents 
 purity in terms of impact parameter in 0-1\% centrality class.
The purity level of $\approx$0.53 (red star on the plot)
is obtained for events with conventional V0M selection,
while best ML classifiers show values $\approx$0.66,
showing increase in purity by about 13\%.
This increase is expected since classifiers use 
information from 5 distinct detecting systems, not 
just from two VZERO scintillators.

\begin{figure}[H]
\sidecaption
\centering
\begin{overpic}[width=0.45\textwidth, bb=0 0 843 560]
{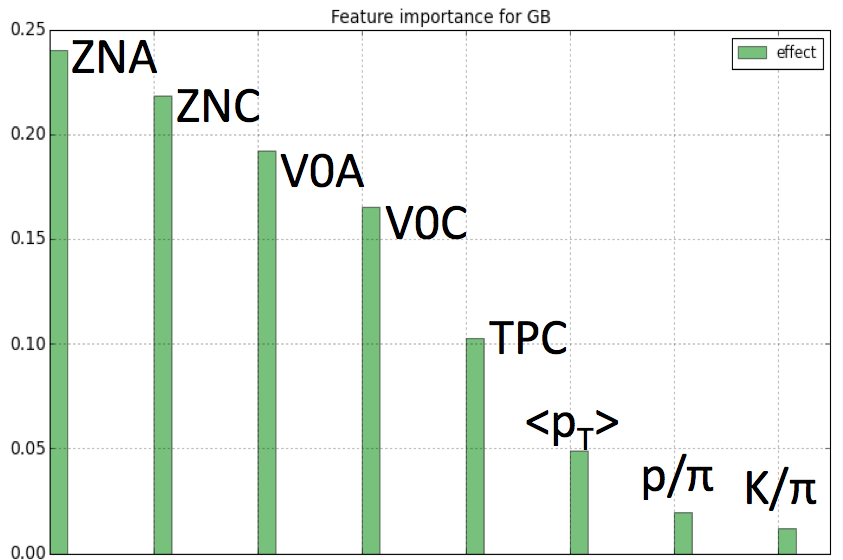}
\end{overpic}
\caption{
Feature importance for BDTG in
classification task for selection of events 
in 0-1\% centrality class.
}
\label{fig:feature_importance_PbPb}
\end{figure}

The same classification task was repeated with an extended set of variables
(three additional features, mentioned in section \ref{sec:ML_task_PbPb},
are event-average transverse momentum $\av\pt$,
K$/\pi$ and p$/\pi$ ratios).
Feature importance histogram is shown in figure \ref{fig:feature_importance_PbPb},
it can be seen that the additional features do not contribute 
to performance improvement much.
Of course, dependence of the results on $\av\pt$
and particle ratios strongly depend the event generator used for the study,
so these features may still be interesting if this approach 
will be applied to real data.

\section{ML task for simulated p-Pb events}

In proton-nucleus collisions, 
observed multiplicities and
other phenomena are much stronger correlated with $N_{\rm part}$
than in Pb-Pb collisions,
while the impact parameter $b$ is only loosely correlated to
$N_{\rm part}$.
Therefore, $N_{\rm part}$
 is a more reliable target for machine-learning classification challenge for p-Pb collisions than $b$. 
Let us split 
\textit{truth} $N_{\rm part}$ distribution and the output from some trained ML estimator
into 5 classes
0-20, 20-40, 40-60, 60-80, 80-100\%
(0-20\% -- for the most central events).
Let us also state ML classification task for p-Pb collisions as follows:
what is the purity of 0-20\% class selected by ML estimator?
In other words, what is the fraction of events, selected by estimator,
which  indeed belong to the truth 0-20\% class of the highest values of $N_{\rm part}$?

\begin{figure}[h]
\centering
\subfigure[a][]
{
\begin{overpic}[width=0.32\textwidth, clip=true, trim=0 0 70 0]
{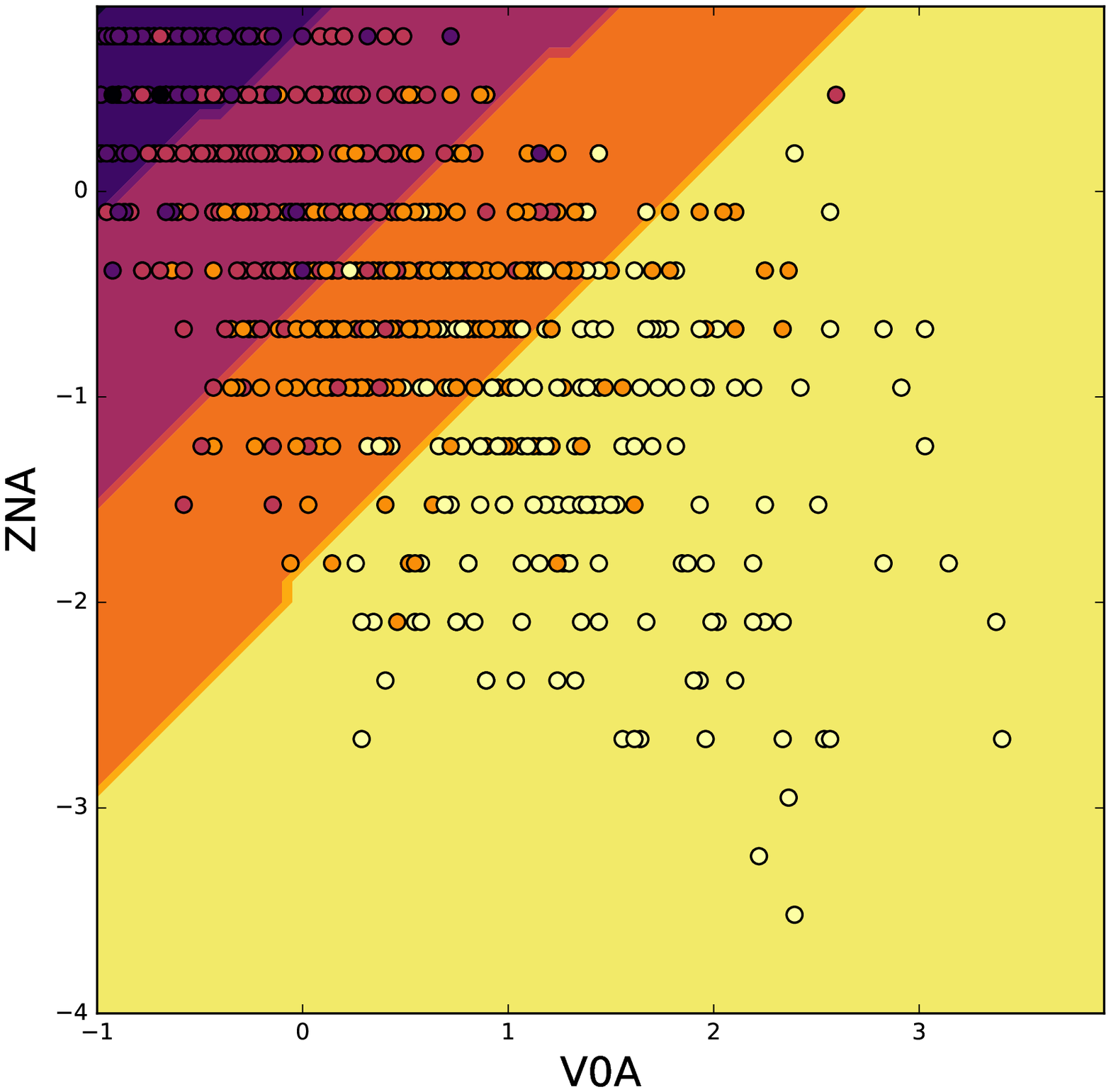}
	\put(7,92){\small V0A, ZNA -- Linear discriminant }
\end{overpic}
}
\subfigure[a][]
{
\begin{overpic}[width=0.32\textwidth, clip=true, trim=0 0 70 0]
{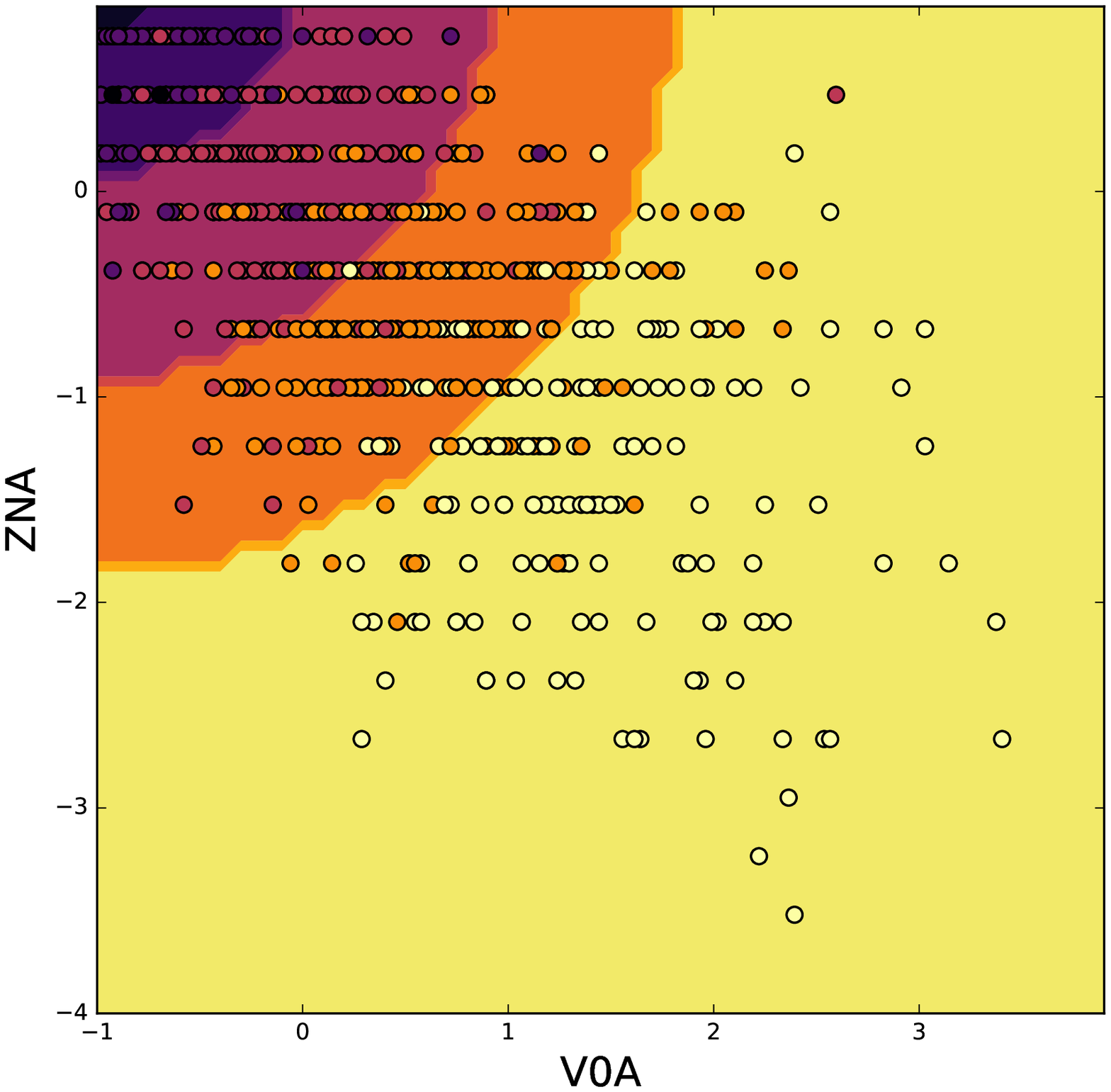}
	\put(7,92){\small V0A, ZNA -- Quadr.  discriminant }
\end{overpic}
}
\subfigure[a][]
{
\begin{overpic}[width=0.32\textwidth, clip=true, trim=0 0 70 0]
{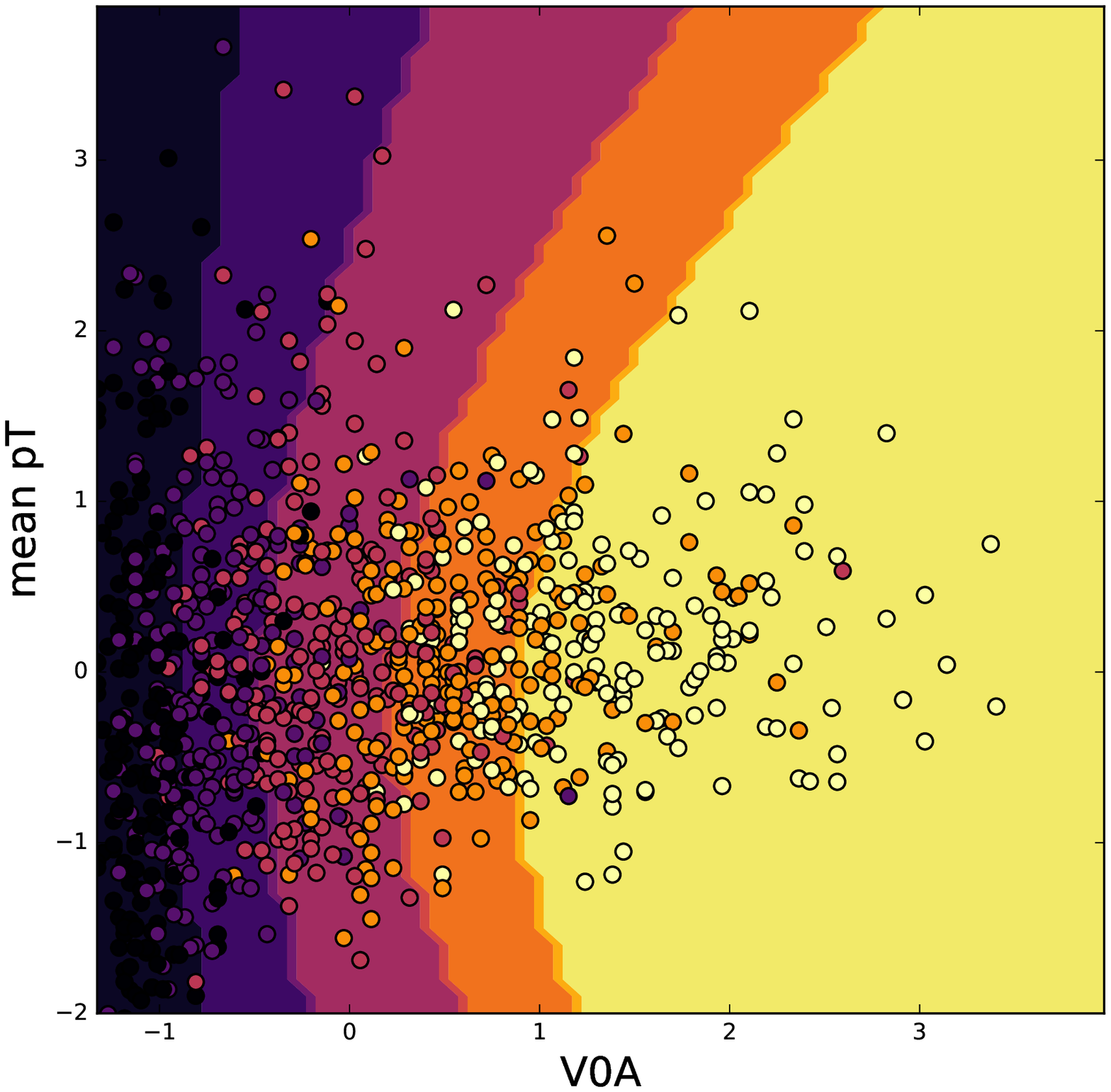}
	\put(7,92){\small V0A, $\av{\pt}$ -- Quadr. discriminant }
\end{overpic}
}
\caption{
2D plots with the following variables on axes:
 V0A and ZNA in pads (a, b) 
and V0A and $\av{\pt}$ in pad (c).
Features are \textit{standardized}
(centered to the mean and 
scaled to unit variance).
Scattered points represent a small fraction of the whole statistics.
Decision boundaries are drawn for five centrality classes
determined by classifiers:  by linear disriminant (a)
and by quadratic discriminant (b, c).
}
\label{fig:decision_boundaries_pPb}
\end{figure}

Scikit-learn package \cite{scikit} was used for this task.
Training and testing were done using 5 mln simulated AMPT events of p-Pb
collisions. 
Again, charged particles for analysis were taken directly from the generator, 
without simulation of the detector response.
At first, only two features were used in the task,
which correspond to ``measurements'' in Pb-fragmentation direction: multiplicity 
in VZERO-A scintillator (denoted here as V0A) and number of neutrons-spectators 
in ZDC-A (denoted as ZNA).
Figure \ref{fig:decision_boundaries_pPb} (a)
is a two-dimensional plot, where scattered points visualize 
 signals in these two detectors (a small part of the whole generated event-sample is shown),
and decision boundaries, obtained after training of linear discriminant classifier,
which drawn by colored areas.
Similar plot is in pad (b) for quadratic discriminant classifier.
Note that decision boundaries between centrality classes
 are not  drawn by hand, as in ZEM-ZDC histogram from figure \ref{fig:V0M_ZEMvsZDC} (b), but instead are calculated as optimal borders between classes.
In the last pad (c) in figure \ref{fig:decision_boundaries_pPb},
the boundaries are shown 
for two another features taken for the classification task
-- multiplicity in VZERO-A
and averaged momentum $\av{\pt}$ of particles in TPC.

\begin{figure}[b]
\centering
\sidecaption
\begin{overpic}[width=0.45\textwidth, ] 
{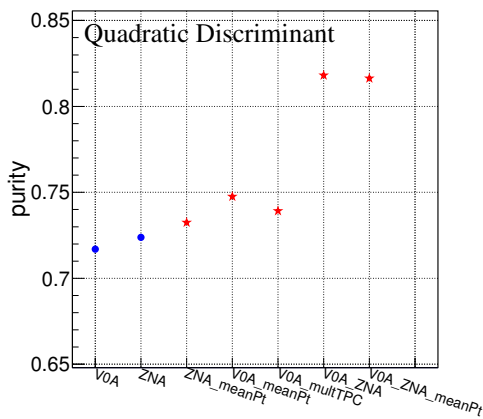}
		\put(16,78){ Quadratic  Discriminant }
\end{overpic}
\caption{
Purity in 0-20\% centrality class
with Quadratic discriminant classifier,
for several combinations of features.
}
\label{fig:purity_pPb}
\end{figure}

\begin{figure}[H]
\centering
\subfigure[a][]
{
\begin{overpic}[width=0.42\textwidth]
{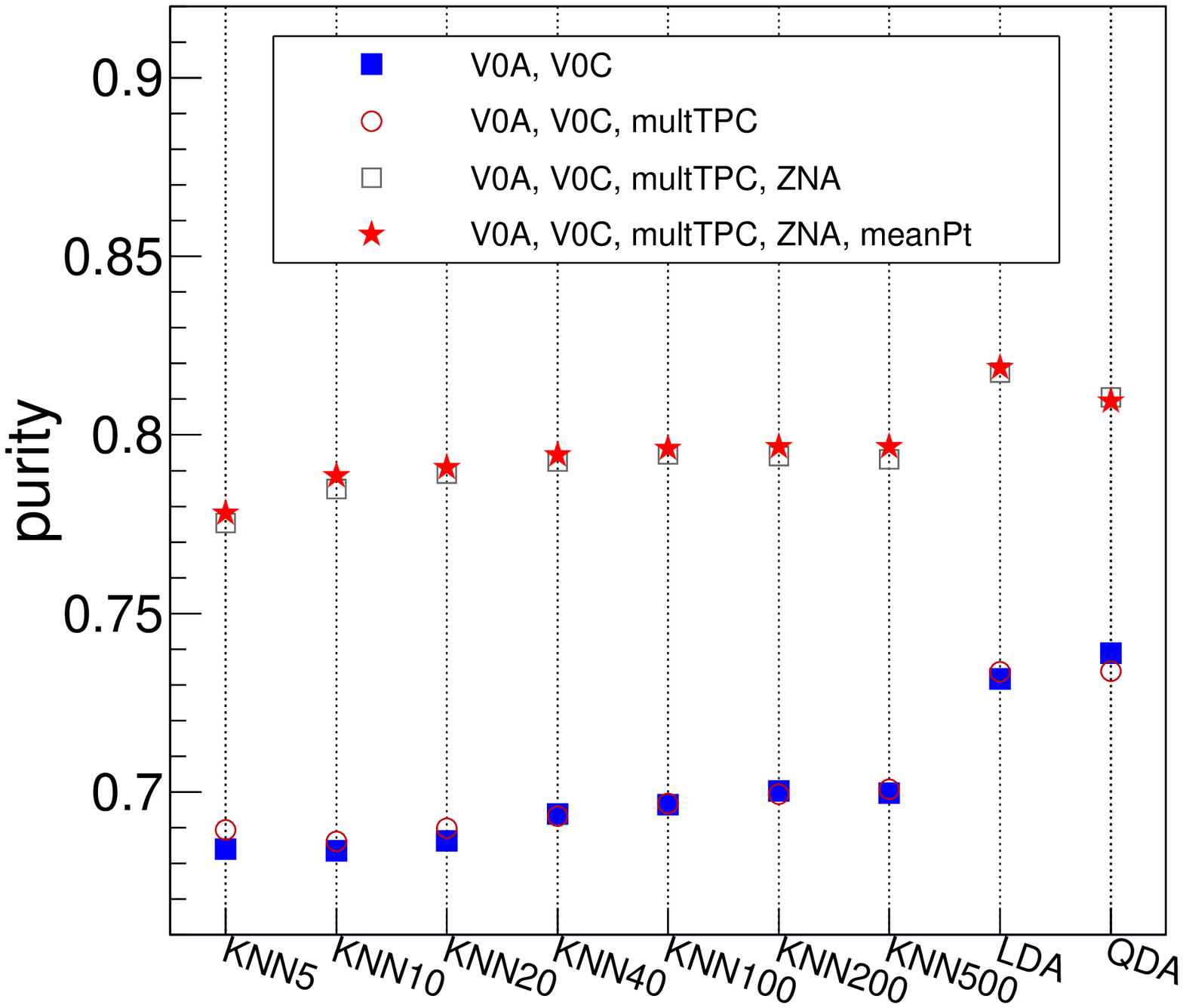}
\end{overpic}
}
\hspace{-0.4cm}
\subfigure[a][]
{
\begin{overpic}[width=0.42\textwidth]
{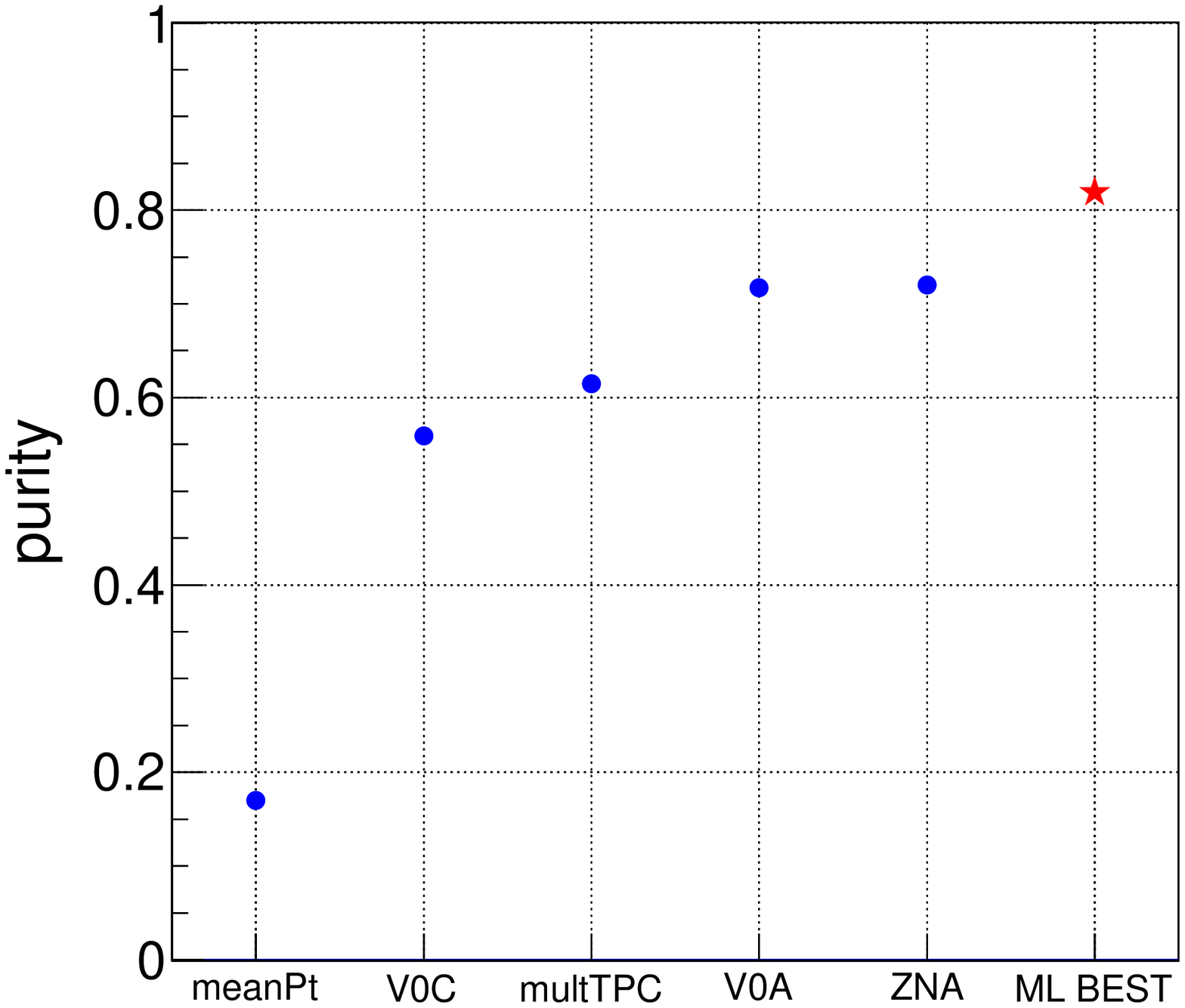}
\end{overpic}
}
\caption{
(a): Purities of 0-20\% centrality class, 
reached with $k$NN, LDA and QDA classifiers
for several feature configurations.
(b): Purities obtained with single-feature classification, in comparison with best ML result with many features.
}
\label{fig:purity_pPb_several_classifiers}
\end{figure}

What are the purities of classification 
with different combination of features?
For figure \ref{fig:purity_pPb},
the quadratic discriminant was taken as a classifier.
Blue circles indicate purity  in 0-20\% class
of events
selected by only one feature: V0A (first point) or ZNA (second point),
giving similar values $\approx$0.72.
For each of the next three points,
one additional feature
was included in ML task, namely,
mean momentum $\av{\pt}$ or multiplicity within TPC.
It can be seen, that both these 
additional features are weak and provide minimal increase in purity, by $\sim$2-3\%. At the same time, when two strong features V0A and ZNA
are combined, purity rises by 10\% to $\approx$0.82.
A similar behaviour of the purity is obtained for several other classifiers
-- $k$ nearest neighbours ($k$NN) with different number of $k$
and for the linear discriminant, see
figure \ref{fig:purity_pPb_several_classifiers} (a).
Pad (b) of this figure shows
purities of 0-20\% class obtained by using 
single-feature classification
(blue circles)
in comparison with the best ML result with many features (red star).

\section{Possible application of ML estimators in NA61/SHINE experiment}

\begin{figure}[H]
\sidecaption
\centering
\begin{overpic}[width=0.21\textwidth, scale=1,bb=0 0 366 369, clip=true, trim=0 0 6 6]
{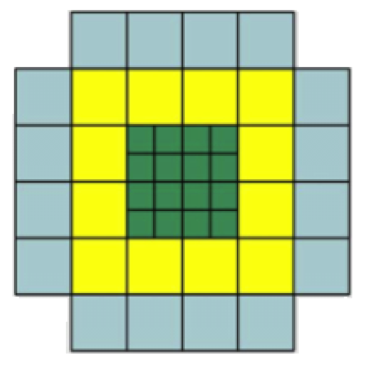}
\end{overpic}
\caption{
Modular structure of the Projectile Spectator Detector 
of the NA61/SHINE facility \cite{NA61_web} .
}
\label{fig:PSD_sketch.png}
\end{figure}

ML-based classifiers can potentially be useful
for centrality estimation in fixed target experiments,
such as NA61/SHINE experiment  at CERN \cite{NA61_web}.
In that experiment, the Projectile Spectator Detector (PSD)
has a modular structure shown in figure \ref{fig:PSD_sketch.png}
and measures the energy of projectile spectators in A-A collisions.
Centrality  is determined by energy in modules of PSD.
Possible improvement can be achieved if the PSD is used in combination with data from several TPC’s,
and 
the machine-learning techniques 
can be applied for that.
Moreover,
it may be useful  to utilize 
energy deposition in the PSD module-by-module, 
and try to benefit from all intrinsic correlations between modules. 
Unfortunately, training of ML estimators requires very robust MC simulations of the PSD.
Situation is complicated 
by the fact that
modules in the PSD are fired not only by spectators, but also by particles born in A-A collision.
Possibly, methods of unsupervised learning could be adopted for this challenge.

\section{Conclusions}

Accurate centrality determination is a baseline for many physics analyses
in heavy-ion collision experiments, 
for example, for fluctuations and correlations studies of A-A and p-A collisions.
In the presented exploratory work,
we tried to 
increase resolution of centrality classes
in terms of the impact parameter $b$ and 
the number of nucleons-participants 
by using machine-learning techniques, which
utilizes signals from several detector subsystems simultaneously.
Compared with conventional centrality estimators,
ML-based estimators allow increasing the resolution and thus the ``purity'' 
of centrality classes
without losses in statistics (i.e., each centrality class contains the same number of events,
but with higher purity).
It could be seen
that improvement in centrality resolution was achieved 
especially for the most central events, 
with increase in purity $\sim$10-13\% compared to conventional centrality selection methods. 

If centrality selection 
is performed using cuts on the signal
from a single 
detecting system,  
results are obviously 
worse 
than if combination of signals from several subdetectors is used. 
At the same time,
simultaneous  usage of just two ``strong'' features may be already enough 
to significantly increase 
the purity of centrality class, 
while  additional ``weak'' features provide very moderate improvement.
ML classifiers allow to draw optimized decision boundaries
between centrality classes, 
and obtained boundaries can have non-trivial shapes. 
It can be especially useful in a multidimensional case
when more than two features are used, which would make  manual parameterization
of decision boundaries very challenging. 

Since the presented work is based on simulations,
efficiency of the ``detectors'' is taken to be unity and contamination by secondary particles is zero,
which makes some features unrealistically strong. For instance,
the  ``ZDC'', used in this study, ideally counts all the neutrons-spectators,
which is not the case at all in a real experiment (real ZDC 
usually has a finite resolution,
it loses some neutrons 
due to combination of spectators into nuclei fragments,
and suffers from contamination by non-spectators).

Additionally, there was no attempt made to incorporate intrinsic correlations 
between selected features, while these correlations are highly
dependent on the physics model of the chosen event generator.
Also,   no attempt was made 
to train and use ML-based estimators for centrality selection in real data, 
since it is not  easy at all
to tune event generator and simulated detector response
 to match data from real detector.
Application of the ML-based centrality estimators 
to real data needs further investigation.

\section{Acknowledgements.}
This work is supported by the Russian Science Foundation, GRANT 16-12-10176.


\end{document}